# Studying the Lunar–Solar Wind Interaction with the SARA Experiment aboard the Indian Lunar Mission Chandrayaan-1

Anil Bhardwaj*, Stas Barabash[†], M. B. Dhanya*, Martin Wieser[†], Futaana Yoshifumi[†], Mats Holmström[†], R. Sridharan*, Peter Wurz[#], Audrey Schaufelberger[#], and Asamura Kazushi[¶]

*Space Physics Laboratory, Vikram Sarabhai Space Centre, Trivandrum, Kerala 695022, India
[†]Swedish Institute of Space Physics, Box 812, 98128, Kiruna, Sweden
[#]Physikalisches Institut, University of Bern, Sidlerstrasse 5,CH-3012, Bern, Switzerland
[¶]Institute of Space and Astronautical Science, 3-1-1 Yoshinodai, Sagamihara, Japan

**Abstract.** The first Indian lunar mission Chandrayaan-1 was launched on 22 October 2008. The Sub-keV Atom Reflecting Analyzer (SARA) instrument onboard Chandrayaan-1 consists of an energetic neutral atom (ENA) imaging mass analyzer called CENA (Chandrayaan-1 Energetic Neutrals Analyzer), and an ion-mass analyzer called SWIM (Solar wind Monitor). CENA performed the first ever experiment to study the solar wind-planetary surface interaction via detection of sputtered neutral atoms and neutralized backscattered solar wind protons in the energy range ~0.01–3.0 keV. SWIM measures solar wind ions, magnetosheath and magnetotail ions, as well as ions scattered from lunar surface in the ~0.01-15 keV energy range. The neutral atom sensor uses conversion of the incoming neutrals to positive ions, which are then analyzed via surface interaction technique. The ion mass analyzer is based on similar principle. This paper presents the SARA instrument and the first results obtained by the SWIM and CENA sensors. SARA observations suggest that about 20% of the incident solar wind protons are backscattered as neutral hydrogen and ~1% as protons from the lunar surface. These findings have important implications for other airless bodies in the solar system.

## INTRODUCTION

The Moon, Earth's natural satellite, is characterized with a surface-bound exosphere and no global magnetic field. Due to the lack of atmosphere and global magnetic field, the solar wind directly impacts on the lunar surface. The solar wind, which consists of around 96% $H^+$, 4% $He^{++}$ and <1% heavier ions [1, 2], can get absorbed on the lunar surface, thermalized, neutralized, and can be backscattered as neutral atoms [3, 4, 5, 6] or it can cause sputtering of atoms from the lunar regolith [7] producing energetic neutral atoms (ENAs). These ENAs carry signature of the surface composition. Thus, by observing ENAs by an orbiting neutral particle instrument and an ion-mass analyzer to measure the solar wind flux, the solar wind–lunar interaction can be investigated in detail.

The sub-keV atom reflecting analyzer (SARA) experiment on the Chandrayaan-1 is meant to study the lunar-solar wind interaction by measuring the ENAs in the lunar environment from a 100 km polar orbit [8, 9]. In the presence of mini-magnetospheres on lunar surface, the solar wind might be hindered to directly impact the lunar surface which would result in a reduction of the ENA flux from these regions. Thus, SARA enables the study of magnetic anomalies present on the lunar surface [8, 10, 11, 12]. Moreover, since the solar wind can penetrate into the permanently shadowed areas (where photons cannot reach) on the lunar surface due to its gyro-radii effect, the imaging of craters near the lunar poles can be done with the SARA. In addition, since the solar wind can access regions beyond the optical terminator, ENA imaging is possible from the region between the optical terminator and solar wind terminator [12].

# THE SARA EXPERIMENT

SARA consists of two sensors, namely CENA (Chandrayaan-1 Energetic Neutrals Analyzer) and SWIM (Solar Wind Monitor), and a DPU (Digital processing Unit). The flight models of the three units are shown in Fig. 1.

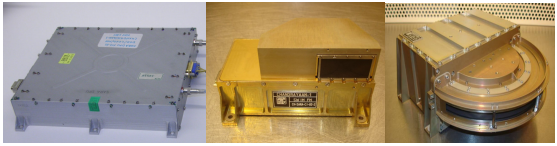

**FIGURE 1.** Flight Models of CENA (right), SWIM (centre) and DPU (left)

CENA detects neutral atoms in the energy range 10 eV – 3.3 keV. The various instrument parameters are described in Table 1. CENA mainly consists of 4 parts which are – a deflection system, a conversion surface, an energy analyzer, and a time of flight (TOF) section. The neutral atoms entering the sensor hit the conversion surface thereby getting converted to positive ions. The ions enter a wave-type energy analyzer that consist of 4 sets of electrodes [13], and by applying suitable voltages at these electrodes, ions of only one particular energy are allowed to pass through. The analyzer also provides efficient UV absorption [8]. Between the conversion surface and the TOF section the ions are post accelerated by 2.4kV. When entering the TOF section, the ions first strike the start surface at grazing incidence and produce secondary electrons that are detected by the start MCP to produce the start pulse.

**Table 1.** CENA and SWIM main characteristics

| Parameter | CENA | SWIM |
|---|---|---|
| Particles to measure | Neutrals | Ions |
| Energy range | 10 eV – 3.3 keV | 10 eV –15 keV |
| Energy resolution | 50 % | 7 % |
| Mass range (amu) | 1 – 56 | 1 – 40 |
| Mass resolution | H, O, Na/Mg/Si/Al-group, K/Ca-group, Fe-group | $H^+$, $He^{++}$, $He^+$, $O^{++}$, $O^+$, > 20 amu |
| Full field-of-view | 15° × 160 ° | 9° × 180° |
| Angular resolution | 9°×25° (E >50 eV) | 4.5° × 22.5° |
| G-factor/sector | $10^{-2}$ cm$^2$ sr eV/eV (at 3.3 keV) 0.2 cm$^2$ sr eV/eV (at 25 eV) | $1.6 \times 10^{-4}$ cm$^2$ sr eV/eV |
| Efficiency (%) | 0.01– 1 | 0.1–5 |
| Sensor mass | 1977 g | 452 g |

After reflection at the start surface, the particles propagate to the stop MCP and produce the stop pulse. The difference in timing of the start and stop pulses gives the TOF. Since the distance the ion traverses between the start surface and stop MCP is known, the velocity of the ion can be deduced. Further, as the energy of the particle is known (from the energy analyzer), the mass of the particle can be obtained. The start MCP has two sets of position sensitive anodes to determine the arrival direction of ions and for exact TOF calculations [8, 9, 13, 14].

SWIM, which is an ion mass analyzer, consists of an electrostatic deflector, energy analyzer, and a TOF section. The electrostatic deflector consists of two plates and by suitably varying the voltage, the direction scanning of the incident ions are achieved. The ions enter an electrostatic energy analyzer, which provides energy analysis, after which the ions are post-accelerated and enter the TOF section. The ions fall on the start surface at grazing incidence producing secondary electrons, which are detected by the start channel electron multipliers (CCEM) to produce the start pulse. The ions themselves get reflected as neutral atoms from the start surface and fall on the stop surface, where they generate again secondary electrons which are collected by the STOP CCEMs to produce the stop pulse. The difference in the timing of the start and stop pulses gives the TOF from which the mass of the ions are deduced in the same manner as described for CENA above. SWIM measures ions in the energy range 10 eV–15 keV. The instrument parameters of SWIM are listed in Table 1, and described in detail in [8, 9, 15].

The digital processing unit (DPU) commands and controls the sensors. It powers the sensors, sets the sensor modes and telemetry modes, acquires data from the sensors, does the time-integration and binning, formats the data, and transfers the data to telemetry. The DPU is built around ADSP21060 DSP processor running at 32 MHz clock frequency and having 4 Mb on-chip memory for operational software.

# OBSERVATIONS AND RESULTS

The field of view (FOV) of CENA is 15° × 160° and that of SWIM is 9° × 180°. They are mounted at 90° to each other on the 3-axis stabilized Chandrayaan-1 top deck, such that CENA is always pointing to the Moon while SWIM is mounted at 90° to the spacecraft velocity vector and the nadir. In this configuration, a fraction of SWIM FOV is always pointed towards the lunar surface. CENA has seven and SWIM has eight viewing directions. The orbital motion of the Chandrayaan-1 is used to scan the

SWIM FOV across the full solar wind angular distribution.

The CENA sensor of SARA has detected the first ENAs from the Moon [16]. These ENAs are mostly hydrogen atoms (H) produced during the interaction of the incident solar wind protons with the lunar regolith. The energy-time spectra of ENAs are shown in the top panel of Fig. 2 for observations made on 6 February 2009. On this day, the orbital plane of Chandrayaan-1 was at an angle of 39° with the day-night terminator. The orbit geometry along with the FOV of CENA and SWIM for this observation is shown in Fig. 3.

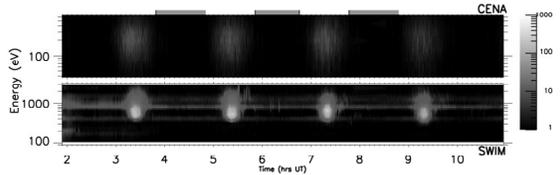

**FIGURE 2.** Raw data showing simultaneous energy-time spectra of hydrogen ENAs reflected from the Moon as measured by CENA (top panel) and the impacting solar wind as measured by SWIM (bottom panel) of the SARA experiment for 4 orbits of Chandrayaan-1 on 6 February 2009. The rectangular grey boxes seen along the X-axis represent the time-intervals when Chandrayaan-1 is on the lunar night-side. The horizontal stripe pattern in the lower panel is background.

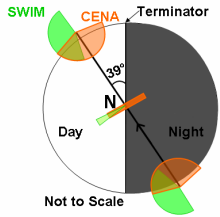

**FIGURE 3.** Field of view of CENA and SWIM at the poles and equator along with the orbit of Chandrayaan-1 for 6 February 2009. Day-night terminator is marked in the figure. CENA is nadir (moon-facing) viewing and SWIM views at 90° to CENA.

The ENAs are seen mostly in the sunlit hemisphere with almost no emission from the night side. There was a simultaneous measurement by SWIM on 6 February 2009. The bottom panel of Fig. 2 show the energy-time spectra of solar wind at the same time when ENAs are seen by the CENA.

The energy spectra of the impinging solar wind ions and the reflected ENAs are shown in Fig. 4. The reflected hydrogen ENA energy spectrum has a broad distribution between 50-200 eV and falls off fast beyond 300 eV, while the energy spectra of incident solar wind peak around 500 eV. This suggests that the reflected hydrogen ENAs loose a significant fraction of the energy of the incident solar wind protons. A good correlation between the reflected energetic neutral flux and the incident solar wind proton flux variation is clearly observed. It is found that ~20% of the incident solar wind protons are backscattered as neutral hydrogen (H atoms) from the lunar surface [16].

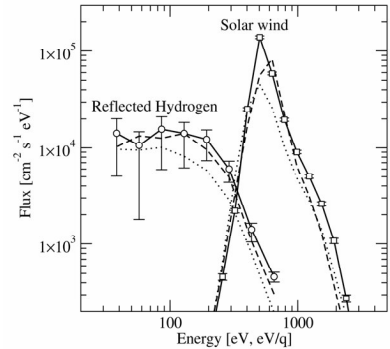

**FIGURE 4.** Fluxes of incident solar wind (square) on the lunar surface and the corresponding reflected H atoms (circle) from the lunar surface as a function of energy, [taken from Ref. 16]. Data corresponding to three orbits out of the 4 orbits shown in Fig. 2.on 6 Feb. 2009 is shown, with dayside equator crossings at 05:22 UTC (solid lines), 07:20 UTC (dashed lines) and 09:18 UTC (dotted lines). Moon was in the magnetosheath, resulting in a broadening of the energy spectrum of the solar wind protons.

In addition to measuring the solar wind ions, SWIM has also detected solar wind ions that are scattered or reflected from the lunar surface. This type of observation is possible due to the mounting of SWIM such that a part of the 180° FOV of SWIM views the lunar surface. Energy-time spectra for each viewing direction of SWIM are shown in Fig. 5 for one such observation on 19 February 2009. The signature of direct solar wind and the reflected ions are separately marked.

The energy spectra of both the solar wind and reflected ions along with the observation geometry are shown in Fig. 6. The energy of reflected solar wind protons is almost the same as that of the incident protons, but the fraction of the reflected protons is ~1% of that of the incident solar wind protons by number. This is consistent with the Kaguya (SELENE) observations [17]. The reflected proton energy spectra are, however, broader compared to that of the incident solar wind.

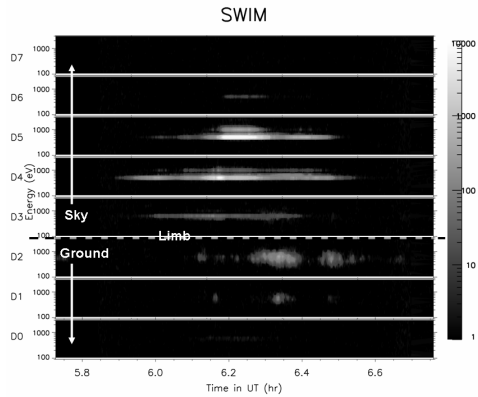

**FIGURE 5.** The energy-time spectra for eight viewing directions measured by SWIM on 19 February 2009, when the Moon was upstream of Earth's bow shock. The limb/lunar horizon is marked in the figure and the viewing direction marked 'Ground' looks at the lunar surface and those marked 'Sky' looks above the horizon.

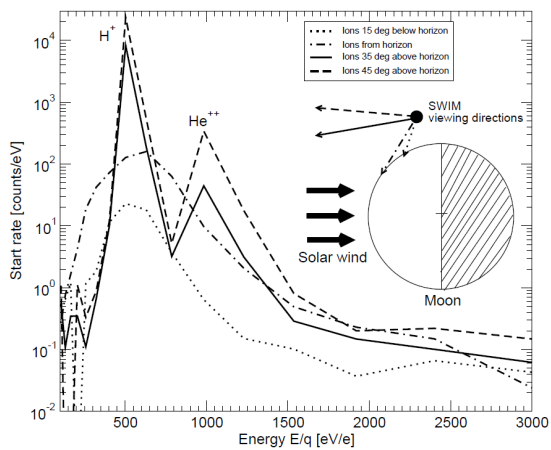

**FIGURE 6.** Energy spectra for SWIM observations on February 19, 2009. On the left, energy spectra for different viewing directions are shown which corresponds to the observation geometry shown on right. Arrows are drawn to indicate the viewing direction of SWIM such that the two upper arrows indicate the viewing direction above horizon and two down-looking arrows indicate the surface viewing directions. The plane of illustration on right side is approximately the equatorial plane.

## DISCUSSION

The SARA observation of ~20% of incident solar wind being reflected as neutral hydrogen atoms has invalidated the earlier assumptions that the entire solar wind impacting the lunar surface is absorbed [18, 19, 20]. The observation of high reflection of incident solar wind protons as ENAs from the Moon suggest that similar processes can be expected on other atmosphere-less bodies in the solar system, such as Mercury, asteroids, Phobos. Also, the observation of the reflection of ~1% of incident solar wind protons from the lunar surface has important implications: These ions can be picked up by the solar wind and accelerated by convective electric field, thus leading to changes in global plasma environment around the Moon [21].

Thus, the SARA observations have opened up a new and exciting area of research on the topic of solar wind-Moon interaction. Our knowledge about the micro-physics of the plasma-surface interaction is quite poor and needs to be explored in detail.

## REFERENCES

1. D. E. Robbins, *J. Geophys. Res.*, 75, 1178-1187 (1970).
2. P. Wurz, *Proceedings of the 11$^{th}$ European Solar physics Meeting- The Dynamic Sun: Challenges for Theory and Observation,* ESA SP- 596, 2005.
3. F. L. Hinton and D. R. Taeusch, *J. Geophys. Res.,* 69, 1341-1347 (1964).
4. F. S. Johnson, *Rev. Geophys. Space Phys.*, 9, 813-823 (1971).
5. R. R. Hodges Jr., *J. Geophys. Res.,* 78, 8055-8064 (1973).
6. R. R. Hodges Jr., *J. Geophys. Res.,85*, 164-170 (1980).
7. P. Wurz, U. Rohner, J. A. Whitby, C. Kolb, H. Lammer, *et al., Icarus,* 191, 486-496 (2007).
8. A. Bhardwaj, S. Barabash, Y. Futaana, Y. Kazama, K. Asamura,*et al.*, *J. Earth Sys. Sci.*, 114, 749-760 (2005).
9. S. Barabash, A. Bhardwaj, M. Wieser, R. Sridharan, T. Kurian, *et al.*, *Current Sci.*, 96, 526-532 (2009).
10. L. L. Hood, *et al.*, *J. Geophys. Res.,* 106, 27825-27839 (2001).
11. J. S. Halekas, D. L. Mitchell, R. P. Lin, S. Frey, L. Hood, *et al.*, *J. Geophys. Res.*, 106,27841-27852 (2008).
12. Y. Futaana, S. Barabash, M. Holmström and A. Bhardwaj, *Planet Space Sci.*, 54, 132-143 (2006).
13. Y. Kazama, S. Barabash, M. Wieser, K. Asamura and P. Wurz, *Planet Space Sci.*, 55, 1518-1529 (2007)
14. Y. Kazama, S. Barabash, A. Bhardwaj, K. Asamura, Y. Futaana, *et al.*, *Adv. Space. Res.*, 37, 38-44 (2006).
15. D. McCann, S. Barabash, H. Nilsson, and A. Bhardwaj, *Planet Space Sci.*, 55, 1190-1196 (2007).
16. M. Wieser, S. Barabash, Y. Futaana, M. Holmström, A. Bhardwaj, *et al. Planet. Space Sci.* (2009), in press, doi: *10.1016/j.pss.2009.09.012*
17. Y. Saito, S. Yokota, T. Tanaka, K. Asamura, M. N. Nishino, *et al.*, *Geophys. Res. Lett.,* 35, L24205 (2008).
18. D. H. Crider and D. R. Taeusch, *J. Geophys. Res.*, 106, 27841-27852 (2008).
19. W. C. Feldman, *et al.*, *J. Geophys. Res.*, 105, 4175-4196 (2000).
20. R. Behrisch and K. Wittmaack, "'Introduction," in *Sputtering by Particle Bombardment III,* edited by R. Behrisch and K. Wittmaack, New York, Springer-Verlag, 1991, pp. 1-13.
21. M. Holmstrom, M. Wieser, S. Barabash, Y. Futaana and A. Bhardwaj, *J. Geophys. Res.* (2009), submitted.